\documentclass{ws-p8-50x6-00}

\begin{document}

\title{Nuclear shadowing and in-medium properties of the $\rho^0$
\footnote{Work supported by DFG and BMBF.}}

\author{T. Falter, S. Leupold and U. Mosel}

\address{Institut fuer Theoretische Physik\\
Universitaet Giessen\\ D-35392 Giessen, Germany}

\maketitle

\abstracts{We explain the early onset of shadowing in nuclear
photoabsorption within a multiple scattering approach and discuss
its relation to in-medium modifications of the $\rho^0$.}

The nuclear photoabsorption cross section is known to be shadowed
at large energies, i.e. $\sigma_{\gamma A}<A\sigma_{\gamma N}$. 
This was at first interpreted as a confirmation of the vector meson dominance 
(VMD) model, which assumes that the photon might fluctuate into vector meson 
states with a probability of order $\alpha_{em}$. To give rise to shadowing, 
these hadronic fluctuations must travel at least a distance $l_V$ that is 
larger than their mean free path inside the nucleus. This so called coherence 
length $l_V$ can be estimated from the uncertainty principle
\begin{equation}\label{eq:coherence}
    l_V\approx\left|k_\gamma-k_V\right|^{-1}
    =\left|k_\gamma-\sqrt{k_\gamma^2-m_V^2}\right|^{-1}
\end{equation}
where $k_\gamma$ and $k_V$ denote the momentum of the photon and the vector 
meson respectively and $m_V$ is the vector meson mass.

Recent photoabsorption data\cite{Bia96,Muc99} indicate an early
onset of shadowing at $E_\gamma\approx$1~GeV. From (\ref{eq:coherence}) one 
sees that the lightest vector meson, e.g. the $\rho^0$, has 
the largest coherence length and therefore its properties determine the onset 
of the shadowing effect. This lead to the interpretation\cite{Bia99} of the
low energy onset of shadowing as a signature of a decreasing $\rho^0$ 
mass in medium since a decrease of $m_\rho$ increases the coherence length 
$l_\rho$. 
%In the following we show that one can also understand the low energy
%onset of shadowing in connection with an increase of the effective $\rho^0$ 
%mass, which is in agreement with dispersion theoretical 
%analyses\cite{Ele97Kon98}.

\begin{figure}[t]
\center
\epsfxsize=18pc % will enlarge or reduce the postscript figures based on the xsize
\epsfbox{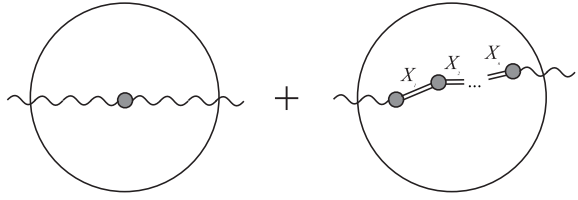} % postscript image file name
\caption{The two amplitudes that contribute to the nuclear forward scattering amplitude in order $\alpha_{em}$. \label{fig:glauber}}
\vspace{-0.5cm}
\end{figure}

A quantitative description of the shadowing effect is possible
within the Glauber model\cite{Bau78}. The nuclear photoabsorption cross 
section can be related via the optical theorem to the nuclear forward 
scattering amplitude. In order $\alpha_{em}$ one finds the two contributions
shown in Fig.~\ref{fig:glauber}. The left amplitude stems from forward
scattering of the photon from one nucleon inside the nucleus.
Summing over all nucleons this amplitude alone leads to the
unshadowed cross section $\sigma_{\gamma A}=A\sigma_{\gamma N}$.
Shadowing arises from the interference with the second amplitude
in order $\alpha_{em}$. Here the photon produces some hadron $X_1$
on one nucleon inside the nucleus. This hadron then scatters
through the nucleus and finally into the outgoing photon which has the same 
momentum and energy as the incoming photon. Since we are dealing with the
forward scattering amplitude and the nucleus has to be in its ground
state after the last scattering event one usually assumes that it stays in its 
ground state during the whole multiple scattering process (multiple scattering 
approximation). One sees that, in principle, shadowing can be explained without
the usage of VMD.

\begin{figure}[t]
  \epsfbox{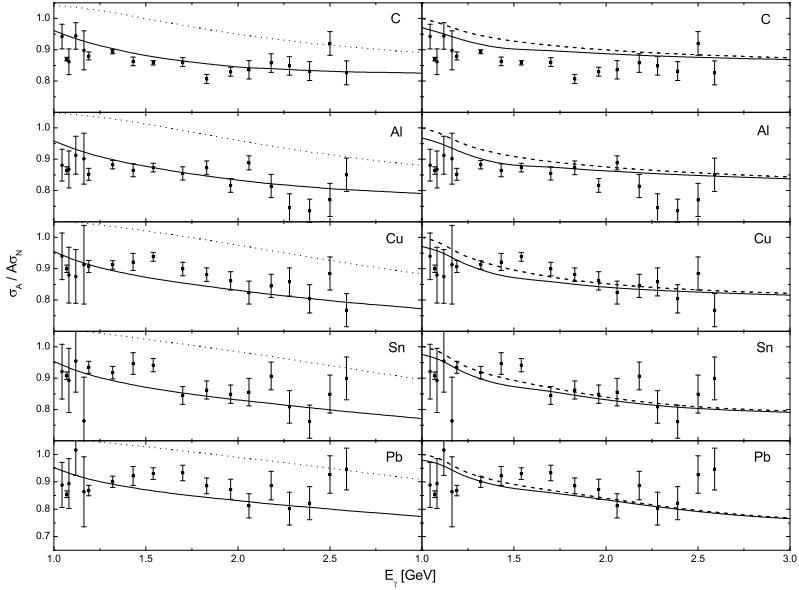} % postscript image file name
  \caption{Calculated ratio $\sigma_{\gamma A}/A\sigma_{\gamma N}$ plotted 
versus the photon energy $E_\gamma$. The {\it left} side shows the result
of the simple Glauber model: with real part of $f_\rho$ (solid lines), 
Re$f_\rho=0$ (dotted lines). The {\it right} side shows the result
of our improved model: contributions from $\rho$, $\omega$ and $\phi$ 
(dashed lines), including the contribution from intermediate $\pi^0$ (solid 
lines). 
\label{fig:results}}
\vspace{-0.5cm}
\end{figure}

In the simple Glauber model one makes use of the eikonal
approximation, assuming that all scattering events at high energies go
predominantly into the forward direction.
This limits the intermediate states $X_i$ to
hadrons which have the quantum numbers of the photon, e.g. the
vector mesons. Neglecting off-diagonal scattering ($VN\rightarrow
V'N$ with $V\neq V'$) and neglecting the widths of the vector
mesons one gets for the total photon nucleus cross section
\begin{eqnarray}\label{eq:cs}
    \sigma_{\gamma A}&=&A\sigma_{\gamma N}
    +\sum_{V=\rho,\omega,\phi}\frac{8\pi^2}{kk_V}\textrm{Re}\biggl\{f_{\gamma V}f_{V\gamma}\int d^2b
    \int_{-\infty}^{\infty}dz_1\int_{z_1}^{\infty}dz_2n(\vec b,z_1)n(\vec b,z_2)\nonumber\\
    & &\quad\times e^{iq_V(z_1-z_2)}\exp\left[-\frac{2\pi}{k_V}f_V\int_{z_1}^{z_2}dz'n(\vec
b,z')\right]\biggr\}.
\end{eqnarray}
Here $n(\vec r)$ denotes the nucleon number density and
$f_{\gamma V}$ and $f_V$ are the vector meson photoproduction and
$VN$ forward scattering amplitudes respectively. In our
calculation\cite{Fal00} we also account for two-body correlations between the
nucleons. In the derivation of (\ref{eq:cs}) one has made
an error of order $A^{-1}$ by summing up infinitely
many multiple scattering terms for the intermediate vector meson.
This is equivalent\cite{Fal00} to the propagation of the vector meson in an 
optical potential, giving rise to an effective in-medium mass and width. The 
momentum transfer $q_V=k-k_V$ in the phase factor arises from 
putting the vector meson on its mass shell. A large momentum transfer $q_V$ 
causes a rapidly oscillating term in the integrand of (\ref{eq:cs}) and 
reduces the shadowing effect. Note that $q_V$ is just the inverse of the 
coherence length $l_V$ as can be seen from (\ref{eq:coherence}).
We now relate the amplitudes $f_{\gamma V}$ and $f_V$ using VMD:
\begin{equation}\label{eq:vmd}
    f_{\gamma V}=f_{V\gamma}=\frac{e}{g_V}f_V.
\end{equation}
The $\rho N$ forward scattering amplitude $f_\rho$ is taken from
dispersion theoretical analyses\cite{Ele97,Kon98}. In the energy region that we
are considering the real part of $f_\rho$ is negative and of the same order of 
magnitude as the imaginary part, leading to an {\em increase} of the effective 
$\rho^0$ mass in medium. Within VMD the negative real part also enters the 
photoproduction amplitude via (\ref{eq:vmd}). This enhances the shadowing
effect and compensates the suppression due to the larger in-medium mass.
In total one gets an increase of shadowing even with a positive mass shift
of the $\rho^0$ in medium. This can be seen from the left side of 
Fig.~\ref{fig:results} where we show our results for the ratio 
$\sigma_{\gamma A}/A\sigma_{\gamma N}$. The calculation that includes the 
negative real part (solid lines) is in perfect agreement with the data. When 
the real part of $f_V$ is neglected, as done in most other calculations, one 
gets the result represented by the dotted curves and clearly underestimates 
the shadowing effect for all nuclei. One even gets anti-shadowing below 2~GeV 
for Pb and 1~GeV for C.

In an improved model\cite{Fal01}, we explicitly sum over multiple 
scattering amplitudes where 1, 2, ... $A$ nucleons participate in the
scattering process. This avoids the error of order $A^{-1}$ that
occurs in the large $A$ limit as described above. We also take the
widths of the vector mesons into account and find that the main
contribution to shadowing at low energies stem from light $\rho^0$
mesons with masses well below the pole mass. These are favored by
the nuclear formfactor because their production is connected with
a small momentum transfer. This is in agreement with our
qualitative understanding of shadowing, since light fluctuations
have a larger coherence length. We also do not hold on to the
eikonal approximation any longer. Dropping this restriction leads to a new 
contribution to the shadowing effect due to $\pi^0$ mesons as intermediate 
states. Since these cannot be produced in the forward direction without
excitation of the nucleus they do not contribute to shadowing at
high energies. In the shadowing onset region, however, they give
rise to 30\% of the total shadowing effect in the case of C and
10\% in the case of Pb as can be seen from the solid lines on
the right side of Fig.~\ref{fig:results}. The dashed lines show the
contribution from intermediate $\rho$, $\omega$ and $\phi$ mesons.
In total one again gets a good description of the shadowing effect.

We have presented a theoretical explanation for the early onset of shadowing
as observed in nuclear photoabsorption. It can be explained by taking the
negative real part of the $\rho N$ forward scattering amplitude into account.
This corresponds to an increase of the effective $\rho^0$ mass in nuclear
medium, in agreement with dispersion theoretical analyses. 
The major contribution to shadowing stems from light $\rho^0$ with masses much
smaller than the pole mass. 
In addition we find contributions from intermediate $\pi^0$ to shadowing in the
onset region.

\end{document}